# The return of (I)DeFiX

Florentina Șoiman [(f)][1,2] [0000-0002-2794-7726], Jean-Guillaume Dumas [1][0000-0002-2591-172X] and Sonia Jimenez-Garces[2]

[1] Univ. Grenoble Alpes, CNRS, LJK, F-38040 Grenoble, France
[2] Univ. Grenoble Alpes, Grenoble INP, CERAG, 38000 Grenoble France
{Firstname.Lastname}@univ-grenoble-alpes.fr

***Abstract***: *Decentralized Finance (DeFi) is a nascent set of financial services, using tokens, smart contracts, and blockchain technology as financial instruments. We investigate four possible drivers of DeFi returns: exposure to cryptocurrency market, the network effect, the investor's attention, and the valuation ratio. As DeFi tokens are distinct from classical cryptocurrencies, we design a new dedicated market index, denoted DeFiX. First, we show that DeFi tokens returns are driven by the investor's attention on technical terms such as "decentralized finance" or "DeFi", and are exposed to their own network variables and cryptocurrency market. We construct a valuation ratio for the DeFi market by dividing the Total Value Locked (TVL) by the Market Capitalization (MC). Our findings do not support the TVL/MC predictive power assumption. Overall, our empirical study shows that the impact of the cryptocurrency market on DeFi returns is stronger than any other considered driver and provides superior explanatory power.*

## I. Introduction

### 1.1. Context

Blockchain was the first tool to bring a decentralized alternative to the existing payment instruments. From that moment on, Fintech and Blockchain technology spread over the whole finance industry and other areas by bringing more automation and innovative solutions. Decentralized finance (DeFi) brings the latest Blockchain-based distributed solutions aiming to provide financial services on a large scale, without intermediaries, using automated protocols. DeFi is a fast-growing sub-sector of the crypto-financial market (Corbet et al., 2021; Ramos & Zanko, 2021). Its solutions cover most functions of the traditional financial system, from funds transfer to margin trading, interest-earning, and market-making. A more detailed description of the DeFi market is provided in appendix section A.

### 1.2. Literature review

Compared to the vast literature on cryptocurrencies, DeFi-related research is scarcer. Gudgeon et al. (2020) assess the interest rates, liquidity, and market efficiency of nowadays main DeFi projects. They find that DeFi tokens (abr. DeFis) are relatively inefficient, while the market is liquid mostly at times of high (platform) utilization. Zetzsche et al. (2020) analyze DeFi platforms' potential, put in the context of the traditional financial economy, and find that DeFis can erode the effectiveness of existing financial regulation. Maouchi et al. (2021) study the bubbles present in the NFT and DeFi markets and discover that COVID-19 and trading volume aggravate the bubble occurrences. They observe that DeFi prices' dynamics are different from pure cryptocurrencies, which distinguishes them from the latter asset class. Moreover, they explain that the total value locked (abr. TVL[1]) should be used as a monitoring tool for the DeFi market (growth), as it indicates the fundamental value of DeFi protocols (Maouchi et al., 2021). Similarly, Corbet et al. (2021) tested for the existence of bubbles and found that DeFi bubbles are mainly self-generated and partially accelerated by ether (ETH) and bitcoin (BTC). At the same time, they test for a possible co-movement between DeFi tokens and cryptocurrencies, which revealed that DeFi tokens represent a separate asset class with some links to conventional cryptocurrencies. In another study, (Harwick & Caton, 2020) argue that while cryptocurrencies have the potential to perform monetary and financial functions superior to those of national currencies, pure decentralized autonomous finance remains an illusory idea as long as it will not integrate real-world identity. Karim et al.(2022) are the first to investigate the risk transmission among non-fungible tokens (NFTs), DeFis, and cryptocurrencies. They show that despite significant risk spillovers in the blockchain markets, namely among DeFis and cryptos, NFTs could greatly serve as an '*investment shield*'. A growing literature debates one of the main drawbacks of the DeFi market, which is a lack of regulation and legal compliance (Anker-Sørensen & Zetzsche, 2021; Aramonte et al., 2021; Chen & Bellavitis, 2019, 2020; Johnson, 2021; Popescu, 2020; Stepanova & Eriņš, 2021; Wronka, 2021).

---

[1] TVL stands for Total Value Locked and refers to the amount of funds attached to a DeFi project.

### *1.2. Motivation and research issue*

DeFi represents a sub-sector of the crypto-financial market (Corbet et al., 2021; Ramos & Zanko, 2021). Motivated by (Corbet et al., 2021; Maouchi et al., 2021; Schär, 2021; Yousaf et al., 2022), who show that DeFi tokens are a distinct asset class compared to conventional cryptocurrencies, our goal is to offer a first analysis of the DeFi market and answer to the following research question: '*What drives the DeFi returns*?'. In their paper, Liu & Tsyvinski (2021) showed that the cryptocurrency market is exposed to the network effect (capturing the user adoption), the momentum effect, and the investor attention but is not impacted by the cryptocurrencies' production factors, macroeconomic factors, or other asset classes (e.g., commodities or stocks). Following their study, we thus investigate four possible drivers of DeFi returns: (1) taking into account that both cryptocurrencies and DeFi tokens run on Blockchain technology and belong to the crypto-market, we assess the cryptocurrencies' return impact on DeFi returns; (2) the exposure to network variables to see if the DeFi market is as well exposed to the same factors as cryptocurrencies (Liu & Tsyvinski, 2021); (3) the investor's attention plays an important role in asset pricing in both cryptocurrency and the stock market, therefore it seems important to assess its impact on the DeFi market as well; and (4) motivated by relevant studies (Ball et al., 2020; Cong et al., 2021; Liu & Tsyvinski, 2021; Pontiff & Schall, 1998) tackling the valuation ratio importance in predicting future returns, we want to investigate if DeFi returns can be predicted by a ratio similar to the 'book-to-market'.

### *1.3. Main findings*

Being an asset class separate from conventional cryptocurrencies (Corbet et al., 2021; Maouchi et al., 2021), we cannot use the CRIX index as a proxy for the DeFi Market. Up to now[2], the whole DeFi market has never been analyzed. Therefore, we contribute to this area of research and design a market index that will enable us to assess the performance of this new asset class and the market as a whole. We compute "DeFiX", a value-weighted market index of all the DeFi tokens. We use the CRIX's index methodology ('an index for Blockchain-based currencies') (Trimborn & Härdle, 2018) and a version of its original code to construct DeFiX, a novel market benchmark for decentralized finance.

In line with Corbet et al. (2021) and Yousaf & Yarovaya (2021), our empirical analysis confirms that the cryptocurrency market strongly influences DeFi returns. DeFi tokens appear to be exposed to their network variables, a result similar to (Liu & Tsyvinski, 2021) for the cryptocurrency market. Our findings reveal that the investor's attention drives DeFi returns on technical terms as "decentralized finance" or "DeFi". Furthermore, we investigated if DeFi returns can be predicted by their corresponding 'book-to-market' ratio. As there is no standard 'book' value for DeFi tokens, we construct a ratio by dividing the Total Value Locked (TVL) by the Market Capitalization (MC). Our findings do not support the TVL/MC predictive power assumption. Out of all the considered drivers, our empirical study shows that the impact of the cryptocurrency market on DeFi returns is the strongest.

The structure of this paper is as follows. The next section exposes the data description and collection. Section 3 introduces the methodology used and data analysis performed, alongside the obtained results. Section 4 comprises the discussion and conclusion.

---

[2] March 2022

## II. Data description and sources

### 2.1. Financial data

We access Coinmarketcap.com and spglobal.com platforms to collect financial data for our DeFi tokens as well as for bitcoin, ether coin, and CRIX index. We download the CRIX index prices from spglobal.com, for which the historical data starts from March 2018. Coinmarketcap is the leading source for financial crypto-related information and has been used as a reliable source by various researchers, including but not limited to (Borri, 2019; Fry & Cheah, 2016; Griffin & Shams, 2020; Grobys & Sapkota, 2019; Howell et al., 2020; Koutmos, 2020; Liu et al., 2022; Stosic et al., 2018; S. Zhang et al., 2019). We extract financial data for 478 DeFi tokens, such as market capitalization, daily prices, and trading volume information. Globally, we have data from 2017 to 2022.

### 2.2. DeFiX

With the financial data extracted from Coinmarketcap, we construct a market index, DeFiX, as the value-weighted return of all underlying tokens. At this step, we are using all the tokens' information, such as price, volume, and market capitalization. Similar to Liu & Tsyvinski (2021)'s approach, we decided to remove the tokens with less than 1,000,000 USD market capitalization. In the construction of DeFiX, we work with daily close prices and use the methodology and original code for the CRIX index (Trimborn & Härdle, 2018). As its name states, CRIX is '*an Index for Blockchain-based currencies*'. Considering that cryptocurrencies and DeFi tokens run on Blockchain technology and belong to the crypto-market, we chose to use the CRIX's index methodology and original code[3] to construct DeFiX. The earliest data for DeFi tokens starts from July 2014 and corresponds to the BitShares platform. After some data cleaning (removing the missing values) and processing, our proposed DeFiX index is composed of 95 tokens and spans from May 2017 to December 2021. A more detailed methodology for our index construction can be found in Appendix B. For other parts of the analysis, we compute as well the weekly and monthly market returns from the daily market returns. A detailed description of all the data used can be found in appendix section C.

### 2.3. Network data

The literature has established that the network effect makes cryptocurrencies more useful as more people join the network / Blockchain (Biais et al., 2020; Cong et al., 2021; Pagnotta & Buraschi, 2018). In other words, this means that the more individuals decide to use BTC, the more valuable the entire Bitcoin ecosystem becomes; hence the price will increase. Liu & Tsyvinski (2021) show that the network effect is one of the cryptocurrencies' price drivers. Therefore, we assess if the same is valid for the DeFi market. We use three primary measures to proxy the network effect on the DeFi market: the number of wallet users, the number of active addresses, and the TVL[4] (total value locked). Similar to (Liu & Tsyvinski, 2021), we retrieve data for the number of transactions and active addresses from Coinmetrics.io. We obtained information for 30 DeFi tokens, out of which we have financial data only for 21. TVL is a unique variable characteristic of the DeFi market. Then, Defillama.com is one of the most complete data aggregators collecting the TVL information for DeFi platforms and financial data for NFTs. We extract the TVL for 503 tokens from Defillama.com, out of which we have financial information (price, market cap. and volume from the coinmarketcap.com) only for 160.

### 2.4. Investor's attention data

Information is at the heart of financial markets and represents an essential element in asset pricing. Just like traditional markets, cryptocurrencies seem to be driven by investor attention as well (Lin, 2020; Liu et al., 2022; Liu & Tsyvinski, 2021; Shen et al., 2019; W. Zhang & Wang, 2020). Therefore, we want to see if the DeFi market's returns have the same driver. We measure the investor's attention with google data (frequency weekly), which is downloaded from the search engine using specific keywords such as 'Decentralized Finance' and 'DeFi'. For further information about data see appendix section C.

The main statistical properties of the variables used in this study are shown in Table 1.

---

[3] From www.quantlet.de

[4] The total value locked refers to the amount of funds committed to a DeFi project. This value is often considered a measure for the platform's success (Maouchi et al., 2021), while at the same time, it could be perceived as a partial substitute to the classical Book value.

Table 1. Summary statistics

| Panel A | | Mean | SD | Max | Min | Skewness | Kurtosis | Durbin-Watson |
|---|---|---|---|---|---|---|---|---|
| Daily | Address Count | 967.763 | 1226.325 | 15259.8 | 0.000 | 2.442 | 12.759 | 0.422 |
| | Transaction Count | 4264.966 | 4417.107 | 19827.385 | 0.000 | 0.96 | 0.017 | 0.008 |
| | TVL | 347010845.408 | 271753418.966 | 1043825445.879 | 34684.038 | 0.452 | -1.251 | 0.001 |
| | Δ Address Count | 0.000 | 0.698 | 4.869 | -4.649 | 0.301 | 2.322 | 2.966 |
| | Δ Transaction Count | -0.001 | 0.509 | 3.897 | -3.332 | 0.306 | 11.733 | 2.869 |
| | Δ TVL | 0.008 | 0.208 | 6.679 | -0.965 | 28.151 | 898.98 | 2.02 |

| Panel B | | Mean | SD | Max | Min | Skewness | Kurtosis | Durbin-Watson |
|---|---|---|---|---|---|---|---|---|
| weekly | "Decentralized finance" search | 21.071 | 24.077 | 100 | 3 | 1.468 | 0.784 | 0.044 |
| | "DeFi" search | 42.035 | 13.793 | 100 | 24 | 1.446 | 1.476 | 0.017 |

| Panel C | | | Mean | SD | Sharpe | Skewness | Kurtosis | Durbin-Watson |
|---|---|---|---|---|---|---|---|---|
| Daily | | DeFiX | 0.125 | 5.725 | 0.022 | -1.173 | 9.563 | 2.083 |
| | | CRIX | 0.118 | 4.589 | 0.026 | -0.422 | 3.356 | 2.016 |
| | | BTC | 0.179 | 4.157 | 0.043 | -0.528 | 8.454 | 2.007 |
| | | ETH | 0.331 | 6.224 | 0.053 | -0.605 | 12.273 | 2.013 |
| Weekly | | DeFiX | 0.871 | 16.045 | 0.054 | -0.295 | 3.004 | 1.921 |
| | | CRIX | 0.588 | 10.733 | 0.055 | -0.954 | 3.655 | 1.821 |
| | | BTC | 1.25 | 10.752 | 0.116 | 0.117 | 2.178 | 1.599 |
| | | ETH | 2.313 | 16.88 | 0.137 | -0.048 | 1.927 | 1.789 |
| Monthly | | DeFiX | 3.79 | 35.591 | 0.106 | 0.402 | 0.646 | 1.975 |
| | | CRIX | 2.529 | 23.784 | 0.106 | -0.008 | -0.958 | 1.694 |
| | | BTC | 5.365 | 26.735 | 0.201 | 1.836 | 9.239 | 1.75 |
| | | ETH | 9.964 | 37.123 | 0.268 | 0.533 | 0.734 | 1.485 |

*This table summarizes the descriptive statistics of the main variables used. In panel A we show the statistical properties of the network variables. For our analyses, we will use as measures the address growth, transaction growth, and TVL growth (marked with 'Δ'). The Durbin-Watson test shows that the growth variables have a lower negative (or positive) autocorrelation than the initial ones. Panel B reports the summary statistics for the two proxies used to measure investor's attention: google searches for "Decentralized finance" and "DeFi".*
*Panel C reports the summary statistics for DeFiX index, CRIX, bitcoin, and ether's returns. Data spans from May 2017 to December 2021.*

Figure 1 plots the cumulative returns of the DeFiX compared to those of the two major cryptocurrencies and the CRIX index. We can observe a strong co-movement among the bitcoin, ether coin, and the DeFi market returns, result confirmed as well by (Corbet et al., 2021). We can see as well that CRIX follows the same trend, confirming the strong co-movement between the cryptocurrency market and the DeFi market.

Figure 1. Cumulative returns of DeFiX against BTC, ETH, and CRIX

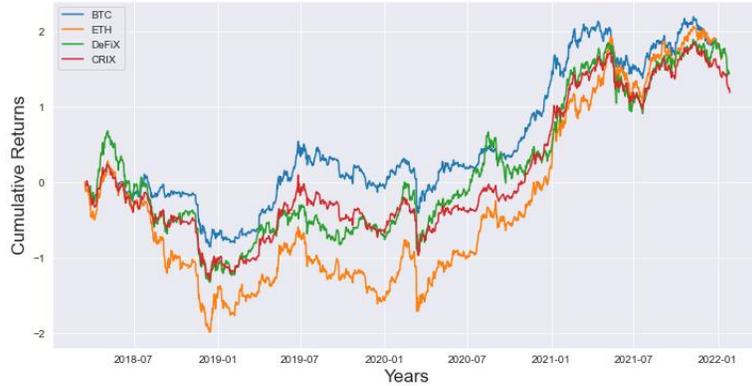

*Cumulative daily returns of the DeFiX (our market index), a DeFi tokens index, against the one of bitcoin (BTC), ether coin (ETH), and CRIX from 2017 to 2022.*

III. Methods and analysis

In this section, we assess three possible drivers for DeFi tokens' returns: the cryptocurrency market, the network effect, and the investor's attention. Furthermore, making a parallel with stocks returns' determinants, we introduce a 'book-to-market ratio' specific to the DeFi market. We further check whether DeFi returns can be predicted by this variable.

**3.1 Exposure to the cryptocurrency market**

Co-movement among securities' returns has always been an important aspect in asset pricing, risk management, and subsequently, asset allocation (Eraker et al., 2003; Oliva & Renò, 2018). Important research has been done on this matter, concerning both the traditional and cryptocurrency markets. Contrasting with the early literature, several studies show that bitcoin is not a dominating coin anymore (Bouri, Shahzad, et al., 2019; Yi et al., 2018), while its strong and persistent influence remains (Bouri et al., 2020; Bouri, Gupta, et al., 2019; Ciaian et al., 2018; Ji et al., 2019; Pereira & Ferreira, 2019; Stosic et al., 2018). Often, connectedness within the crypto-market is observed among the leading cryptocurrencies only, and it is stronger in the short run (Corbet et al., 2019; Yarovaya & Zięba, 2022).

We investigate the predictive power of cryptocurrencies (BTC, ETH, and CRIX index) for the DeFi market returns. For addressing such an issue, we begin by proposing a new index of the DeFi market, namely, the DeFiX index. We then analyze the impact of cryptocurrencies' returns on the DeFiX index and on 15 of the leading DeFis[5]. The rationale behind our selection for BTC and ETH cryptocurrencies is the following: Ethereum is the main technology used in developing DeFi platforms (Popescu, 2020; Ramos & Zanko, 2021), while BTC and ETH are the main digital coins used as leverage in DeFi operations (Aramonte et al., 2021; Schär, 2021). CRIX is the index for the cryptocurrency market, and since the current literature (Corbet et al., 2021; Karim et al., 2022; Maouchi et al., 2021) shows an existing relationship between bitcoin and certain DeFi tokens, it seems relevant to assess whether the crypto-market as a whole has an impact on the DeFi market.

Before carrying out a regression analysis, we study the correlation between cryptocurrencies and DeFis. Table 2 reports a strong correlation between cryptocurrency and the DeFi market. To avoid possible multicollinearity issues, we thus further assess the exposure of the DeFi market to the cryptocurrency market by making individual checks instead of regressions with all the variables together.

Table 2. Pearson correlation check between the cryptocurrency market and the DeFi market

|       | DeFiX | Rt_ETH | Rt_BTC | CRIX |
|-------|-------|--------|--------|------|
| **DeFiX** | 1 |        |        |      |

---

[5] The selection choice for the biggest DeFi tokens have been made based on their average market capitalization over the past 100 days.

|  | | | | |
|---|---|---|---|---|
| Rt_ETH | 0.758*** | 1 | | |
| Rt_BTC | 0.619*** | 0.692*** | 1 | |
| CRIX | 0.537*** | 0.606*** | 0.655*** | 1 |

*The data frequency is weekly. By looking at our results, we can observe that there is a strong correlation between cryptocurrency and the DeFi market. \*\*\* denotes significance levels based on the respective p-value (\*:10%, \*\*:5%, and \*\*\*:1%).*

### 3.1.1 The predictive power of bitcoin

We begin the analysis of the DeFi exposure to the cryptocurrency market by first checking the power of influence of bitcoin. We regress the DeFiX index returns and the 15 main DeFi tokens' returns (LUNA, AVAX, WBTC, UNI, DAI, LINK, FTM, XTZ, AAVE, GRT, MKR, CAKE, RUNE, CRV, and LRC) against bitcoin returns on weeks t-1 and t-2. The results of our regressions are shown in Table 3.

$$(1)\ Rt\ DeFiX = \alpha + \beta_1\ R_{t-1}BTC + \beta_2\ R_{t-2}BTC + \varepsilon,$$
$$(2)\ Rt\ Major\ DeFi\ token = \alpha + \beta_1\ R_{t-1}BTC + \beta_2\ R_{t-2}BTC + \varepsilon.$$

Table 3: The predicting power of bitcoin on DeFiX and DeFi Tokens.

|  | Constant | Rt-1 BTC | Rt-2 BTC | $R^2$ |
|---|---|---|---|---|
| Rt DeFiX | -0.0014 (-0.165) | **0.8710*** (6.066)** | -0.1867 (-1.302) | 26.7% |
| Rt LUNA | 0.0099 (0.999) | **0.3018*** (1.749)** | 0.0258 (0.150) | 4.9% |
| Rt AVAX | 0.0078 (0.975) | 0.1551 (1.120) | -0.0055 (-0.040) | 1.7% |
| Rt WBTC | 0.0035 (0.892) | **0.3888*** (5.734)** | 0.0016 (0.024) | 32.7% |
| Rt DAI | 0.000 (0.140) | **-0.0187*** (-1.789)** | 0.0088 (0.843) | 1.9% |
| Rt LINK | 0.0076 (0.684) | **1.0342*** (5.332)** | **-0.3841*** (-1.982)** | 17.2% |
| Rt UNI | 0.0026 (0.391) | 0.1517 (1.333) | -0.0229 (-0.202) | 1.9% |
| Rt FTM | 0.0113 (0.355) | **0.7775*** (3.656)** | -0.3229 (-1.520) | 8.2% |
| Rt XTZ | -0.0062 (-0.566) | **0.5633*** (2.964)** | 0.0153 (0.081) | 11.9% |
| Rt AAVE | 0.0202 (1.084) | 0.04231 (1.307) | -0.3341 (-1.033) | 0.7% |
| Rt GRT | 0.0022 (0.226) | 0.1281 (0.756) | 0.0687 (0.412) | 1.8% |
| Rt CAKE | 0.0049 (0.525) | 0.1962 (1.204) | -0.0212 (-0.130) | 1.7% |
| Rt MKR | -0.0017 (-0.203) | **0.9517*** (6.418)** | **-0.3380*** (-2.282)** | 23.7% |
| Rt RUNE | 0.0178 (1.599) | **0.3202*** (1.660)** | -0.1135 (-0.589) | 2% |
| Rt CRV | -0.0070 (-0.747) | 0.1225 (0.757) | 0.0250 (0.155) | 1.2% |
| Rt LRC | -0.0004 (-0.029) | **0.7513*** (3.361)** | -0.0938 (-0.420) | 11.7% |

*\*\*\* denotes significance levels based on the respective p-value (\*:10%, \*\*:5%, and \*\*\*:1%). The standard t-statistic value is shown in parentheses. Here we check if BTC can help to predict DeFi tokens' returns. $R_{t-1}$ ($R_{t-2}$) refers to one (two) week(s) lagged BTC returns. $R_t$ stands for weekly return. The number of observations for each regression is 265.*

We find that one-week lagged BTC returns have a positive and significant impact on future DeFiX returns, as well as on several leading tokens. LUNA, WBTC, MKR, LINK, FTM, XTZ, RUNE, and LRC seem to be positively influenced by one-week lagged BTC returns. This result is in line with (Corbet et al., 2021). Out of the fifteen major DeFi tokens included in our assessment, nine show to be impacted by BTC returns. If most of our results show a positive impact of the previous week's BTC returns on DeFi returns, we observe that DAI returns are negatively and significantly impacted by BTC's one week ahead returns. LINK and MKR seem to be as well negatively impacted by BTC two weeks ahead return, which means that an increase in BTC will result in a return decrease for LINK, MKR, and DAI. Based on these obtained results, we argue that, in addition to the BTC's influence arising from its role in DeFi operations, DeFi platforms might as well benefit from the cryptocurrencies' hype and, hence, are driven by BTC variations.

### 3.1.2 The predictive power of ether coin

Celeste et al. (2020) suggest that there is a relationship between ETH success and the growth of Ethereum based platforms. As most DeFi platforms are built on Ethereum technology, we check the predictive power of ETH for DeFi's returns. Therefore, we regress the DeFiX index returns and the main DeFi tokens against ETH one-week and two-weeks lagged returns. We summarize our results in Table 4.

$$(3)\ Rt\ DeFiX = \alpha + \beta_1\ R_{t-1}ETH + \beta_2\ R_{t-2}ETH + \varepsilon,$$

$$(4)\ Rt\ Major\ DeFi\ tokens = \alpha + \beta_1\ R_{t-1}ETH + \beta_2\ R_{t-2}ETH + \varepsilon,$$

Table 4: The predicting power of ether coin on DeFiX and DeFi tokens returns.

|  | Constant | Rt-1 ETH | Rt-2 ETH | $R^2$ |
|---|---|---|---|---|
| *Rt DeFiX* | -0.0050 | 0.5835*** | 0.0254 | 36.4% |
|  | (-0.653) | (5.869) | (0.259) |  |
| *Rt LUNA* | 0.0093 | 0.2489* | -0.0096 | 5.2% |
|  | (0.930) | (1.942) | (-0.076) |  |
| *Rt AVAX* | 0.0072 | 0.0644 | 0.0610 | 2.1% |
|  | (0.897) | (0.627) | (0.602) |  |
| *Rt WBTC* | 0.0043 | 0.2138*** | 0.0002 | 18.7% |
|  | (0990) | (3.860) | (0.004) |  |
| *Rt DAI* | 0 | -0.0092 | 0.0034 | 1% |
|  | (0.119) | (-1.182) | (0.441) |  |
| *Rt LINK* | 0.0047 | 0.07827*** | -0.2298* | 21.5% |
|  | (0.426) | (5.576) | (-1.660) |  |
| *Rt UNI* | 0.0016 | 0.1158 | 0.0131 | 3.4% |
|  | (0.239) | (1.380) | (0.158) |  |
| *Rt FTM* | 0.0088 | 0.6356*** | -0.2301 | 11.4% |
|  | (0.731) | (4.092) | (-1.502) |  |
| *Rt XTZ* | -0.0090 | 0.4262** | 0.0729 | 16.5% |
|  | (-0.842) | (3.098) | (0.537) |  |
| *Rt AAVE* | 0.0196 | 0.3087 | -0.2218 | 0.7% |
|  | (1.047) | (1.283) | (-0.934) |  |
| *Rt GRT* | 0.0031 | 0.0488 | 0.0373 | 0.7% |
|  | (0.312) | (0.385) | (0.299) |  |
| *Rt CAKE* | 0.0041 | 0.1999* | -0.0527 | 2.5% |
|  | (0.442) | (1.656) | (-0.443) |  |
| *Rt MKR* | -0.0047 | 0.7467*** | -0.2205* | 30.8% |
|  | (-0.572) | (7.111) | (-2.129) |  |
| *Rt RUNE* | 0.0156 | 0.2341 | 0.0010 | 4.1% |
|  | (1.412) | (1.650) | (0.007) |  |
| *Rt CRV* | -0.0087 | 0.0452 | 0.1300 | 3.1% |
|  | (-0.939) | (0.380) | 1.107 |  |
| *Rt LRC* | -0.0038 | 0.6271*** | -0.0491 | 16.8% |
|  | (-0.305) | (3.889) | (-0.309) |  |

*\*\*\* denotes significance levels based on the respective p-value (\*:10%, \*\*:5%, and \*\*\*:1%). The standard t-statistic value is shown in parentheses. Here we check if ETH can help to predict DeFi tokens' returns. $R_{t-1}$ ($R_{t-2}$) refers to one (two) week(s) lagged ETH returns. $R_t$ stands for weekly return. The number of observations for each regression is 265.*

ETH predicting power for DeFis' returns is similar to the one already reported for BTC. Out of the fifteen major DeFi platforms included in our test, eight show a significant increase in their returns due to the ETH returns increase the week before. We find that one-week lagged ETH returns have an important impact on future DeFiX returns, and several leading tokens. Similar to BTC predicting power assessment, MKR and LINK seem to be strongly positively (negatively) influenced by one (two) past weeks ETH returns; This result is in line with the existing literature (Agosto & Cafferata, 2020; Celeste et al., 2020; Chang & Shi, 2020), which show that compared to other pure financial assets like BTC or LTC, Ethereum is the main technology used in the development of service-based Blockchain instruments and distribution of tokens. Premised on this fact, Ethereum is an important contributor in the tokens' pricing as well. LUNA and CAKE show a weak return impact by one-week lagged ETH. As we can observe, for the rest of the sample (AVAX, DAI, UNI, AAVE, GRT, RUNE, CRV), ETH does not show any significant predicting power.

### 3.1.3 The predictive power of CRIX

At this point, we assess whether the crypto-market as a whole has an impact on DeFis' returns. We regress the DeFiX index returns and the main DeFi tokens' returns over the one-week and two-weeks lagged CRIX returns. Results are presented in Table 5.

$$(5)\ Rt\ DeFiX = \alpha + \beta_1\ R_{t-1} CRIX + \beta_2\ R_{t-2} CRIX + \varepsilon,$$

$$(6)\ Rt\ Major\ DeFi\ tokens = \alpha + \beta_1\ R_{t-1} CRIX + \beta_2\ R_{t-2} CRIX + \varepsilon,$$

Table 5: The predicting power of CRIX on DeFiX and DeFi tokens returns.

|  | Constant | Rt-1 CRIX | Rt-2 CRIX | $R^2$ |
|---|---|---|---|---|
| **Rt DeFiX** | 0.0040 (0.480) | 0.9714*** (5.317) | -0.1011 (-0.586) | 26.6% |
| **Rt LUNA** | 0.0116 (1.213) | 0.4888** (2.304) | 0.1184 (0.591) | 11.2% |
| **Rt AVAX** | 0.0081 (1.048) | 0.3210* (1.871) | 0.0462 (0.285) | 6.6% |
| **Rt WBTC** | 0.0060* (1.828) | 0.6729*** (9.282) | -0.0719 (-1.051) | 52.4% |
| **Rt DAI** | 0 (0.037) | -0.0220* (-1.659) | 0.0064 (0.510) | 2.4% |
| **Rt LINK** | 0.0121 (1.114) | 1.3039*** (5.392) | -0.3743 (-1.639) | 20.4% |
| **Rt UNI** | 0.0028 (0.443) | 0.3020** (2.150) | 0.0210 (0.158) | 7.6% |
| **Rt FTM** | 0.0130 (1.142) | 1.3576*** (5.354) | -0.4257* (-1.778) | 19.3% |
| **Rt XTZ** | -0.0020 (-0.190) | 0.6800** (2.869) | 0.1313 (0587) | 15.3% |
| **Rt AAVE** | 0.0202 (1.089) | 0.5281 (1.284) | -0.2615 (-0.674) | 0.9% |
| **Rt GRT** | 0.0030 (0.316) | 0.4649** (2.194) | -0.0774 (-0.387) | 5.2% |
| **Rt CAKE** | 0.0053 (0.581) | 0.5677*** (2.812) | -0.1525 (-0.800) | 6.8% |
| **Rt MKR** | 0.0021 (0.262) | 1.2922*** (7.369) | -0.3257* (-1.968) | 34% |
| **Rt RUNE** | 0.0181* (1.684) | 0.5152** (2.159) | 0.0201 (0.089) | 7.3% |
| **Rt CRV** | -0.0069 (-0.765) | 0.3759* (1.874) | 0.0347 (0,183) | 6.1% |
| **Rt LRC** | 0.0038 (0.308) | 1.1481*** (4.197) | -0.1279 (0.495) | 18.2% |

*\*\*\* denotes significance levels based on the respective p-value (\*:10%, \*\*:5%, and \*\*\*:1%). The standard t-statistic value is shown in parentheses. Here we check if CRIX can help to predict DeFi tokens' returns. $R_{t-1}$ ($R_{t-2}$) refers to one (two) week(s) lagged CRIX returns. $R_t$ stands for weekly return. The number of observations for each regression is 265.*

While assessing the CRIX predicting power for DeFis', we observe that almost all leading DeFi tokens and the market index, DeFiX, are positively and significantly impacted by the increase in CRIX returns the week before. This result could be justified by a combined effect / predictive power of all cryptocurrencies applied through the CRIX index. In spite of being developed on Ethereum technology, AAVE is the only token from our sample that seems to not be significantly influenced by the crypto-market at all. As in the assessment of BTC influence, the DAI token seems to be negatively impacted by one week before CRIX returns. Similarly, FTM and MKR are negatively impacted by two-weeks lagged CRIX returns, meaning that an increase in CRIX's returns from two weeks before will generate a decrease in MKR and FTM returns. AVAX and CRV show a small impact in their returns due to CRIX returns increase the week before. We also find out that FTM and MKR returns are influenced by the CRIX return values one and two weeks before.

In this section, we have investigated the predictive power of cryptocurrencies (BTC, ETH, and CRIX index) on the DeFiX market. Our findings show that BTC and ETH have a comparable strong and significant influence in predicting DeFi returns. While assessing the CRIX predicting power over DeFis, we observe that almost all leading DeFi tokens and the market index, DeFiX, are positively and significantly impacted. This result could be justified by a combined effect / predictive power of all cryptocurrencies applied through the CRIX index. Overall, our empirical analysis shows that the impact of the cryptocurrency market on DeFi returns is strong and statistically significant.

## 3.2 Network effect

"Whether real or virtual, networks have a fundamental economic characteristic: the value of connecting to a network depends on the number of other people already connected to it." (p.174, Shapiro & Varian, 1999). Katz & Shapiro (1985, 1986) have made some of the first contributions to the network economy literature, underlying for the first time the relationship between the fundamental value[6] and network effects. This issue becomes more complex when the network is virtual, for example, in the case of software businesses, where "*the linkages between nodes are invisible*" but not inexistent (p.174, Shapiro & Varian, 1999). The (virtual) network effect exists in the crypto-market as well, and multiple papers assess its impact on the returns of cryptocurrencies (Biais et al., 2020; Cong et al., 2021; Pagnotta & Buraschi, 2018). The conclusion is quite clear: network effects make cryptocurrencies more useful as more people join the (Blockchain) network, and as a result of this, the entire crypto-ecosystem becomes more valuable. Considering that cryptocurrencies and DeFi tokens both run on Blockchain technology and both belong to the same crypto-market, it seems relevant to assess whether the network effect also exists in the DeFi market.

A significant part of the literature on cryptocurrencies has shown the importance of network factors in the valuation of cryptocurrencies (Bhambhwani et al., 2019; Biais et al., 2020; Liu & Tsyvinski, 2021; Sockin & Xiong, 2020). To the best of our knowledge, there is no prior work exploring the network effect for DeFi tokens valuation.

Maouchi et al. (2021) and Saengchote (2021) argue that TVL is a reliable tool for DeFi market monitoring: "*TVL is already considered by the crypto-community as one of the main indicators of DeFi markets size and growth*" (Maouchi et al., 2021, page 7). One of (Cong et al., 2021)'s main assumptions with regards to token valuation is based on the expected growth of the network as a result of the platform's increased productivity. Accordingly, we recognize TVL as an important variable in the network assessment. The rationale behind this represents simply the fact that TVL represents the amount of funds committed to the DeFi business. This means that the more people join and transact on DeFi platforms (translating in more funds and bigger TVL), the larger the network will be.

Active addresses and transactions count are commonly used as network variables in the crypto-market assessment (Hinzen et al., 2022; Koutmos, 2020; Liu & Tsyvinski, 2021; Nadler & Guo, 2020). We thus choose three factors as proxies for the DeFi network effect: the active address count and the transactions count (like for the crypto-asset market), as well as the total value locked (TVL). As far as we know, we are the first to consider TVL as a proxy for network effect. To measure the network growth, we are using: the address growth, the transaction growth, and the TVL growth. The growth (noted Δ) of network variables is computed using the logarithmic difference method.

Table 6: Pearson correlation check among the network variables

|  | Δtransac | Δaddress | Δtvl |
|---|---|---|---|
| Δ transac | 1 | | |
| Δ address | 0.385*** | 1 | |
| Δ tvl | -0.019 | 0.056 | 1 |

*This table shows the correlation matrix of the network variables. The network factors include address growth (Δ address), transaction growth (Δ transac), and TVL growth (Δ tvl). \*\*\* denotes significance levels based on the respective p-value (\*:10%, \*\*:5%, and \*\*\*:1%).*

We first assess the correlation between the three network variables we consider. The results presented in Table 6, show a positive relationship between transaction and address growth, with a correlation coefficient of 0.385. The TVL growth seems not correlated with any of the other two network variables. To evaluate the exposure of Defi returns to network variables, we run regressions of the DeFix returns over the growth of transactions, the growth of addresses, and the growth of TVL. Rt DeFiX represents the weekly returns of our index for the DeFi market:

$$(7) \quad Rt\ DeFiX = \alpha + \beta_1\ \Delta Transac + \varepsilon,$$

---
[6] fundamental value refers to the estimation of a security (e.g., stock, bonds, cryptocurrency) value, considering its incurred risks and the information within the market. It may be different from the market value, which is the current price. We cannot accurately determine the *real* value of a security, this is why we try to estimate it.

$$(8) \quad Rt\ DeFiX = \alpha + \beta_1\ \Delta\ Address + \varepsilon,$$
$$(9) \quad Rt\ DeFiX = \alpha + \beta_1\ \Delta\ TVL + \varepsilon,$$
$$(10) \quad Rt\ DeFiX = \alpha + \beta_1\ \Delta\ Transac + \beta_2\ \Delta\ Address + \beta_3\ \Delta\ TVL + \varepsilon.$$

The above regressions assess the direct exposure of DeFiX returns to the growth of network variables. The last regression regroups all network variables as independent variables in order to consider possible interaction effects. We present the regressions' results in Table 7.

Table 7: Rt DeFiX exposure to network variables

|  | (7) | (8) | (9) | (10) |
|---|---|---|---|---|
| Δ Trans | 0.049*** (2.745) |  |  | 0.044** (2.199) |
| Δ Address |  | 0.021* (1.89) |  | 0.009 (0.744) |
| Δ TVL |  |  | 0.095** (2.145) | 0.095** (2.168) |
| $R^2$ | 2.80% | 1.30% | 1.90% | 5.10% |

*Table 4 reports the exposure of DeFiX's returns to the (averaged) network factors. \*\*\* denotes significance levels based on the respective p-value (\*:10%, \*\*:5%, and \*\*\*:1%). The standard t-statistic value is shown in parentheses. The data frequency is weekly. For each regression, we had 239 observations.*

We find that DeFiX returns are exposed to each network variable. The coefficients for transaction count and TVL growth are strongly significant in regressions (7), (9), and (10). This result confirms that the performance of the DeFi market, represented here by the DeFiX index, is driven by its fundamental value (as proxied by the network variables). In finance, fundamental value refers to the 'real' risk-adjusted value of a security, which may be different from the market value. As for stocks, where the fundamental value can be derived from accounting information on the firm's operation (Brainard et al., 1990), we believe that DeFis derive their value from the network. This rationale is based on (Katz & Shapiro, 1986) theory, which debates that the technology's adoption and value evolution are linked to its network size. Out of the three variables tested, our results show that the transaction count and the TVL growth have the strongest relationship with DeFi returns. These two variables reflect the platform's success (value), therefore, an increase in transaction count and TVL translates into an increase in financial returns for DeFi tokens holders. On the same point but in the crypto-market, (Liu & Tsyvinski, 2021) find that all network variables play an important role in the valuation of cryptocurrencies.

Furthermore, we assess the impact of network variables' growth on individual tokens' returns. We perform panel regressions using the network data we have. We first regress the DeFis' returns on the network variables change for all DeFis' tokens for which we have network information. This allows us to assess as many tokens as possible and thus to enlarge the picture from only leading DeFis. Afterward, we make a panel regression using only the leading tokens, in line with the previous section.

### 3.2.1. Exposure to TVL growth

We start with our first network variable: TVL growth. We have TVL information for 160 tokens. Therefore, we first perform the regression of the DeFi tokens returns over the change in TVL for all 160 tokens. We then only focus our attention on the 15 leading tokens. We lagged TVL growth from 1 week up to 4 weeks. Results are shown in Table 8 and Table 9.

$$(11) \quad Rt\ DeFis_i(t) = \alpha_i(t) + \sum_{l=1}^{4} \beta_l, i\ \Delta TVL_i(t-l) + \varepsilon_i(t).$$

Table 8: Panel OLS, returns of 160 DeFi tokens against their lagged TVL growth

|  | Parameter | Std. Err. | T-stat | P-value | Lower CI | Upper CI |
|---|---|---|---|---|---|---|
| const | -1.5875*** | 0.2419 | -6.5621 | 0.0000 | -2.0617 | -1.1133 |
| Δ TVL(t-1) | 6.2969*** | 0.7362 | 8.5532 | 0.0000 | 4.8538 | 7.7401 |
| Δ TVL(t-2) | 1.3949 | 0.9946 | 1.4024 | 0.1608 | -0.5549 | 3.3447 |
| Δ TVL(t-3) | 2.3299** | 1.0588 | 2.2005 | 0.0278 | 0.2544 | 4.4054 |
| Δ TVL(t-4) | 1.5552 | 1.0485 | 1.4833 | 0.138 | -0.5001 | 3.6105 |

*R2 is 1.42%, R2 between is 9.43%. We regressed the returns of 160 DeFi tokens against their Δ TVL lagged from 1 week to 4 weeks. Time effects included. We used weekly returns and weekly TVL growth. For this regression, we had 7453 observations. \*\*\* denotes significance levels based on the respective p-value (\*:10%, \*\*:5%, and \*\*\*:1%).*

Our results show that 1 and 3 weeks ahead TVL growth can positively predict DeFis' return increase.

We then perform the same panel regression, but this time we consider only the 15 major DeFi tokens. This approach was used in section 3.1, when we assessed the DeFis' exposure to the cryptocurrency market.

Table 9: Panel OLS, returns of 15 leading DeFi tokens against their lagged TVL growth

|  | Parameter | Std. Err. | T-stat | P-value | Lower CI | Upper CI |
|---|---|---|---|---|---|---|
| const | 0.9281 | 0.7787 | 1.1918 | 0.2337 | -0.6005 | 2.4567 |
| Δ TVL(t-1) | **8.8468\*\*\*** | **3.2887** | **2.6901** | **0.0073** | **2.391** | **15.303** |
| Δ TVL(t-2) | -0.1661 | 3.7732 | -0.044 | 0.9649 | -7.573 | 7.2408 |
| Δ TVL(t-3) | -0.0873 | 3.9044 | -0.0224 | 0.9822 | -7.7517 | 7.5771 |
| Δ TVL(t-4) | 2.8899 | 4.3563 | 0.6634 | 0.5073 | -5.6617 | 11.441 |

*R2 is 1.17%, R2 between is 16.54%. We regressed the returns of major DeFi tokens against their Δ TVL lagged from 1 week to 4 weeks. Time effects included. We used weekly returns and weekly TVL growth. For this regression, we had 942 observations. \*\*\* denotes significance levels based on the respective p-value (\*:10%, \*\*:5%, and \*\*\*:1%).*

While assessing the TVL growth predicting power for the leading DeFis' returns, our results confirm the previous findings. More specifically, we observe that all leading tokens are positively and significantly impacted by the 1-week ahead increase in TVL growth. As mentioned before, TVL represents the amount locked in DeFi platforms. Our results indicate that when the value locked in the platform increases, the value (return) of the corresponding DeFi token increases as well.

### 3.2.2. Exposure to transaction count growth

Our second network variable is transaction count growth. We have transaction count information for 21 tokens. First, we regress all 21 tokens returns against their lagged transaction growth. In a second step, we refer only to the 15 leading tokens. We lagged transaction growth from 1 week up to 4 weeks. Results are shown in Table 10 and Table 11.

$$(12) \; Rt \; DeFis_i(t) = \alpha_i(t) + \sum_{l=1}^{4} \beta_l, i \; \Delta trans_i(t-l) + \varepsilon_i(t).$$

Table 10: Panel OLS, returns of 21 DeFis against their lagged Transaction growth

|  | Parameter | Std. Err. | T-stat | P-value | Lower CI | Upper CI |
|---|---|---|---|---|---|---|
| const | 0.4434\*\*\* | 0.169 | 2.6233 | 0.0087 | 0.112 | 0.7748 |
| Δ Trans(t-1) | -2.9266\*\*\* | 0.441 | -6.6368 | 0.0000 | -3.7911 | -2.0621 |
| Δ Trans(t-2) | -1.0038\*\* | 0.5032 | -1.9948 | 0.0461 | -1.9904 | -0.0173 |
| Δ Trans(t-3) | -0.657 | 0.5054 | -1.2999 | 0.1937 | -1.6477 | 0.3338 |
| Δ Trans(t-4) | -1.4298\*\*\* | 0.4765 | -3.0008 | 0.0027 | -2.3638 | -0.4957 |

*R2 is 0.94%, R2 between is -9.72%. We regressed the returns of 21 DeFi tokens against their Δ Transac. lagged from 1 week to 4 weeks. Time effects included. We used weekly returns and weekly Transaction growth. For this regression, we had 5407 observations. \*\*\* denotes significance levels based on the respective p-value (\*:10%, \*\*:5%, and \*\*\*:1%). The standard t-statistic value is shown in parentheses.*

Table 11: Panel OLS, returns of 15 leading DeFis against their lagged Transaction growth

|  | Parameter | Std. Err. | T-stat | P-value | Lower CI | Upper CI |
|---|---|---|---|---|---|---|
| const | 0.573 | 0.3587 | 1.5971 | 0.1104 | -0.1307 | 1.2766 |
| Δ Trans(t-1) | -1.6691 | 1.067 | -1.5643 | 0.1179 | -3.7619 | 0.4238 |
| Δ Trans(t-2) | 0.1894 | 1.1481 | 0.165 | 0.869 | -2.0626 | 2.4415 |

| | Parameter | Std. Err. | T-stat | P-value | Lower CI | Upper CI |
|---|---|---|---|---|---|---|
| Δ Trans(t-3) | 0.3374 | 1.143 | 0.2952 | 0.7679 | -1.9046 | 2.5795 |
| Δ Trans(t-4) | -0.3854 | 1.0655 | -0.3617 | 0.7176 | -2.4753 | 1.7046 |

*R2 is 0.22%, R2 between is -0.07%. We regressed the returns of major DeFi tokens against their Δ Transac. lagged from 1 week to 4 weeks. Time effects included. We used weekly returns and weekly Transaction growth. For this regression, we had 1832 observations. The standard t-statistic value is shown in parentheses.*

According to the first regression (21 tokens), DeFi returns appear to be negatively and significantly impacted by the change in one week, two weeks, and four weeks ahead transaction growth. This indicates that the more transactions are placed on the DeFi platform, the lower the return of the corresponding token. While a bit surprising, these results could be explained by the increased volatility and large financial bubbles present in the DeFi market (Maouchi et al., 2021). At the same time, (Hau et al., 2021) show that cryptocurrency returns can be positively (negatively) influenced by transaction activity if the market state is bullish (bearish). The same could be true for the DeFi market, as the presence of large bubbles justifies the less efficient prices. While out of the scope of this paper, more investigations on this matter are necessary.

The regression performed with only leading tokens does not report any significant results. This means that no major tokens considered in this analysis are impacted by transaction growth.

### 3.2.3. Exposure to address count growth

Our last network variable is the address count growth. We have address count information for 21 tokens. Therefore, we regress all 21 tokens' returns against their address growth in a panel data regression analysis. In a second step, we then refer only to the leading tokens. We lagged address growth from 1 week up to 4 weeks. Results are shown in Table 12 and Table 13.

$$(13)\ Rt\ DeFis_i(t) = \alpha_i(t) + \sum_{l=1}^{4} \beta_{l,i} \Delta address_i(t-l) + \varepsilon_i(t).$$

Table 12: Panel OLS, returns of 21 DeFis against their lagged Address growth

| | Parameter | Std. Err. | T-stat | P-value | Lower CI | Upper CI |
|---|---|---|---|---|---|---|
| const | **0.4498*** | **0.169** | **2.6618** | **0.0078** | **0.1185** | **0.7812** |
| Δ Address(t-1) | **-1.8314*** | **0.424** | **-4.3192** | **0.000** | **-2.6626** | **-1.0001** |
| Δ Address(t-2) | -0.512 | 0.5001 | -1.0238 | 0.306 | -1.4925 | 0.4684 |
| Δ Address(t-3) | -0.7134 | 0.5048 | -1.4132 | 0.1576 | -1.703 | 0.2762 |
| Δ Address(t-4) | -0.6625 | 0.4554 | -1.4546 | 0.1458 | -1.5553 | 0.2304 |

*R2 is 0.43%, R2 between is 1.42%. We regressed the returns of 21 DeFi tokens against their Δ Address lagged from 1 week to 4 weeks. Time effects included. We used weekly returns and weekly Address growth. For this regression, we had 5432 observations. \*\*\* denotes significance levels based on the respective p-value (\*:10%, \*\*:5%, and \*\*\*:1%). The standard t-statistic value is shown in parentheses.*

Table 13: Panel OLS, returns of 15 leading DeFis against their lagged Address growth

| | Parameter | Std. Err. | T-stat | P-value | Lower CI | Upper CI |
|---|---|---|---|---|---|---|
| const | 0.5424 | 0.359 | 1.511 | 0.131 | -0.1617 | 1.2466 |
| Δ Address(t-1) | -0.5388 | 0.7926 | -0.6798 | 0.4967 | -2.0935 | 1.0159 |
| Δ Address(t-2) | 0.5559 | 0.9693 | 0.5735 | 0.5664 | -1.3455 | 2.4572 |
| Δ Address(t-3) | 0.8046 | 0.9787 | 0.8221 | 0.4111 | -1.1151 | 2.7243 |
| Δ Address(t-4) | 0.4923 | 0.8255 | 0.5964 | 0.551 | -1.1269 | 2.1116 |

*R2 is 0.11%, R2 between is -1.71%. We regressed the returns of major DeFi tokens against their Δ Address lagged from 1 week to 4 weeks. Time effects included. We used weekly returns and weekly Address growth. For this regression, we had 1836 observations. The standard t-statistic value is shown in parentheses.*

Similar to our previous results concerning transaction growth, we can observe that in the first panel data regression (21 tokens), DeFi returns are negatively impacted by the change in one week ahead address growth. The regression performed with only the leading tokens does not report any significant results. This translates as, the more addresses are created for a specific DeFi platform, the lower the return of the corresponding token in the next week. Motivated by the findings from (Hau et al., 2021) that include address number in their sample, we think that DeFi returns can be positively (negatively) influenced by address growth if the market state is bullish (bearish). These results confirm the inefficiency of DeFi prices, in line with (Corbet et al., 2021; Maouchi et al., 2021), although further investigation would be necessary.

Same as our findings from section 3.2.2, the regression performed with only leading tokens does not report any significant results. This means that no major tokens considered in this analysis are impacted by address growth.

In this section, we have assessed the predictive power of network variables (the active address count, the transactions count, and TVL) on the DeFi market returns. Our results show that the DeFi returns are strongly influenced by their network variables, like cryptocurrencies (Liu & Tsyvinski, 2021). Overall, our empirical study shows that the impact of the TVL growth over DeFi returns is stronger than any other network variable considered and provides superior explanatory power.

### 3.3 Investor attention

Investor attention impacts financial market characteristics, including liquidity, returns, and volatility. This hypothesis has been proved to be valid in both traditional markets and cryptocurrencies ones (Andrei & Hasler, 2015; Ciaian et al., 2015; Lin, 2020; Liu et al., 2022). Often, investor attention is associated with the under- and overreaction of investors, being considered a key determinant of potential market mispricing (Andrei & Hasler, 2015; Baker & Wurgler, 2007; Shiller, 2000). Tversky & Kahneman (1974) investigate individuals' capacity to make rational decisions in situations of uncertainty. They show that people tend to perceive certain events in a specific way that ignores the laws of probability. In the financial markets, this could translate in situations such as investors seeing that the cryptocurrencies' price keeps increasing, they will consider it a growing market and invest in, ignoring the laws of probability such as this being just a bubble phase (J. Li & Yu, 2012). The traditional asset-pricing theory assumes that markets assimilate new information rapidly, and securities' prices adjust and incorporate the news. Shiller (1980) proposes the use of volatility measure in assessing market efficiency. By looking at the high volatility present in the cryptocurrency market (M. Hu et al., 2018; Noda, 2021; Tran & Leirvik, 2020), we assume that their prices cannot be linked to economic realities but rather to investors' irrational decision making described by psychological models such as the one developed by (Tversky & Kahneman, 1974). This claim is based on the premise that the crypto-market could be inefficient (prices deviate from their fundamental value), a subject highly debated in the literature (Y. Hu et al., 2019; Kristoufek & Vosvrda, 2019; Tran & Leirvik, 2019; Wei, 2018; Yaya et al., 2021) and which brought mixed conclusions.

Given the vast literature introducing the importance of investor attention in the cryptocurrency market and its key role in asset pricing (Corbet et al., 2019; Guégan & Renault, 2021; Guzmán et al., 2020; Ibikunle et al., 2020; Z. Li et al., 2021; Lin, 2020; Liu & Tsyvinski, 2021; Nasir et al., 2019; W. Zhang & Wang, 2020), it seems relevant to study this phenomenon on the DeFi market.

Similar to the (Liu & Tsyvinski, 2021) paper, we build proxies for investor's attention with Google searches. Hence, to assess the impact of investor's attention on DeFi returns, we extracted the number of google searches for DeFi related keywords: 'Decentralized finance' (noted "dcfin" in our following table and equation) and 'DeFi'. We selected these two keywords while thinking about the unique features of the DeFi market. The name, 'DeFi" is an abbreviation for decentralized finance. While words like "decentralization" and "decentralized finance" are generally associated with Blockchain and cryptocurrencies, DeFi platforms are the only technology that makes complete finance decentralization possible.

The second key-word represents simply the general name for all the tokens belonging to DeFi platforms. In this way, we assume that the investors who have some knowledge about Blockchain will know about 'decentralized finance', but only the connoisseurs (informed investors) will be the ones understanding the "DeFi" term.

The interest of an investor for a particular investment might not be immediate. For that reason, we perform our tests on different periods, and we lag the Google search-based proxy from one up to three weeks (between the investor's interest and his potential investment). Consequently, we perform the following regressions:

$$(14)\ Rt\ DeFiX(t) = \alpha + \sum_{i=1}^{3}\beta_i\ dcfin(t-i) + \sum_{j=1}^{3}\beta_{j+3}\ DeFi(t-j) + \varepsilon,$$

$$(15)\ Rt\ DeFis_i(t) = \alpha + \sum_{i=1}^{3}\beta_i\ dcfin(t-i) + \sum_{j=1}^{3}\beta_{j+3}\ DeFi(t-j) + \varepsilon.$$

Results are shown in Table 14.

Table 14: Predicting power of investor's attention

|  | Constant | dcfin (t-1) | dcfin (t-2) | dcfin (t-3) | DeFi (t-1) | DeFi (t-2) | DeFi (t-3) | $R^2$ |
|---|---|---|---|---|---|---|---|---|
| Rt DeFiX | 0.0085 (0.834) | 0.0430 (1.110) | -0.0456 (-1.087) | **-0.0796* (-2.035)** | -0.0083 (-0.091) | 0.0540 (0.565) | **0.2063** (2.277)** | 4.3% |
| Rt LUNA | **0.0181* (1.659)** | 0.0477 (1.149) | 0.0176 (0.394) | 0.0029 (0.069) | 0.0193 (0.198) | -0.0955 (-0.931) | 0.0223 (0.228) | 1.30% |
| Rt AVAX | 0.0125 (1.461) | 0.0291 (0.895) | 0.0024 (0.068) | -0.0274 (-0.835) | 0.015 (0.196) | -0.0327 (-0.408) | 0.0226 (0.296) | 1% |
| Rt WBTC | **0.0117** (2.304)** | 0.0176 (0.914) | 0.0291 (1.404) | 0.0121 (0.622) | -0.0371 (-0.818) | -0.0444 (-0.932) | -0.0348 (-0.768) | 1.10% |
| Rt DAI | -0.0001 (-0.138) | -0.0012 (-0.495) | -0.0009 (-0.334) | 0.0005 (0.215) | -0.0009 (-0.159) | 0.0027 (0.438) | -0.0027 (-0.462) | 0.50% |
| Rt LINK | 0.0172 (1.33) | 0.0086 (0.175) | -0.0128 (-0.241) | **-0.1046** (-2.105)** | 0.049 (0.423) | 0.046 (0.378) | **0.3053*** (2.636)** | 4% |
| Rt UNI | 0.0064 (0.946) | 0.0226 (0.873) | 0.0089 (0.320) | -0.0203 (-0.775) | 0.0578 (0.948) | -0.0461 (-0.720) | -0.0021 (-0.034) | 1.80% |
| Rt FTM | 0.0166 (1.231) | 0.0537 (1.045) | 0.0232 (0.421) | -0.0716 (-1.383) | -0.1109 (-0.920) | -0.1269 (-1.002) | 0.1316 (1.090) | 2.60% |
| Rt XTZ | -0.0046 (-0.372) | -0.0083 (-0.177) | **-0.1218** (-2.417)** | **-0.1389*** (-2.936)** | -0.0647 (-0.587) | 0.0234 (0.202) | 0.0833 (0.755) | 4.80% |
| Rt AAVE | 0.0217 (1.072) | -0.031 (-0.402) | -0.0475 (-0.575) | 0.0096 (0.123) | 0.0093 (0.051) | 0.0718 (0.378) | -0.0223 (-0.123) | 0.30% |
| Rt GRT | 0.0086 (0.813) | 0.0263 (0.657) | 0.0297 (0.690) | -0.0235 (-0.582) | 0.1067 (1.133) | 0.06 (0.607) | -0.0416 (-0.442) | 1.80% |
| Rt CAKE | 0.0084 (0.829) | 0.001 (0.026) | -0.0111 (-0.268) | -0.0205 (-0.527) | 0.0099 (0.110) | -0.0455 (-0.477) | -0.0647 (-0.712) | 0.50% |
| Rt MKR | 0.0071 (0.690) | 0.0492 (1.257) | -0.0412 (-0.98) | **-0.085** (-2.153)** | 0.0534 (0.581) | -0.0151 (-0.156) | **0.2321** (2.523)** | 5.90% |
| Rt RUNE | **0.0246** (2.073)** | 0.0713 (1.580) | -0.0272 (-0.561) | -0.0274 (-0.602) | 0.0871 (0.821) | 0.0034 (0.03) | 0.0422 (0.398) | 2.60% |
| Rt CRV | -0.0006 (-0.065) | 0.0292 (0.787) | 0.0392 (0.982) | 0.0108 (0.287) | 0.0753 (0.863) | 0.056 (0.611) | 0.0586 (0.671) | 1.40% |
| Rt LRC | 0.012 (0.831) | 0.0773 (1.409) | -0.0177 (-0.301) | **-0.101* (-1.824)** | 0.05 (0.388) | -0.0527 (-0.389) | **0.2285* (1.771)** | 4.20% |

*Here we assess the predictive power of investor's attention, proxied by google search terms "Decentralized finance" (denoted as "dcfin") and "DeFi". We regressed DeFiX index, and major DeFi tokens' returns against the google search lagged from one up to three-weeks horizons. Rt stands for weekly returns. For each regression, we had 250 observations. *** denotes significance levels based on the respective p-value (*:10%, **:5%, and ***:1%). The standard t-statistic value is shown in parentheses.*

We find that Google searches have a significant impact just on some DeFi token returns. Furthermore, the impact of the search for 'decentralized finance' seems stronger than for 'DeFi'. Decentralized finance is a concept that is not limited to DeFi platforms, but it has been around for a longer period of time (e.g. (Dillinger & Fay, 1999)). Moreover, the idea of decentralization has been widely discussed since Blockchain's invention (Chuen & Deng, 2018; Guo & Liang, 2016; Harwick & Caton, 2020; Karaulova, 2017; Leonhard, 2019; Swan, 2017; Yeoh, 2017), therefore this result is not surprising taking into account that DeFi is based on Blockchain technology.

We cannot ignore the fact that 'decentralized finance' search has a negative impact on the DeFi market, meaning that an increased investor attention (proxied by 'decentralized finance') will decrease DeFi returns. This could be explained by the high volatility present and insufficient development of the young DeFi market. Moreover, since 'decentralized finance' is not limited only to DeFi and as we have shown in section 3.1, the DeFi market is exposed to the crypto-market, our results could be justified by possible linkage and transmissions from one asset to another. More research is needed on this matter. 'DeFi' term is specific to DeFi markets only. Hence, our findings showing a positive impact prove that an increase in investor attention (proxied by 'DeFi' searches) induces an increase in tokens' returns.

By looking at our results, we can observe that only four[7] out of fifteen major tokens seem to be impacted by google searches. Since globally, the market returns (represented by DeFiX) can be predicted from investor's attention, we therefore assume that smaller tokens might be influenced by google searches. Liu & Tsyvinski (2021) and Shen et al., (2019) report an increased investor attention during the period of high volatility. Similarly, (Karim et al., 2022) paper reveals an increased investor attention on DeFi market following the enhanced volatility. As our results do not provide enough evidence to draw any conclusions regarding DeFi market efficiency, we believe that more research is necessary.

### 3.4 DeFi valuation ratio

In this section, we investigate the ability of a DeFi market specific 'book-to-market' ratio to predict DeFi returns.

While researchers and practitioners have always tried to identify the variables that could help to predict future returns, the motivation to use the book-to-market ratio as a possible driver has arisen after the findings of Fama & French (1992). In their paper (Fama & French, 1992), Fama and French reveal that the book-to-market ratio has more explanatory power for the cross-sectional variations in stock returns than traditional risk measures. On another perspective, Lakonishok et al. (1994) and La Porta (1996) believe that the book-to-market ratio is just an evidence of mispricing, as investors form biased expectations concerning the future prospects and firm value based on accounting information. Many papers have already shown the ability of the book-to-market ratio to predict future returns on both traditional markets (Ball et al., 2020; Pontiff & Schall, 1998) and the cryptocurrency market (Cong et al., 2021). Liu & Tsyvinski (2021) do not find any significant relationship between the valuation ratio and the future cryptocurrency return. Motivated by the existing literature on the cryptocurrency market and by the studies of other financial markets, we find it relevant to investigate whether the book-to-market ratio can predict DeFi returns.

The classic book-to-market ratio used in the stock market compares a company's book value to its market value. The book value refers to the accounting data (total value of assets minus total liabilities), and the market value represents the market price of one share multiplied by the total number of shares outstanding.

If the market value is easily available for both stocks and cryptocurrencies, we know that there is no direct measure of the book value for cryptocurrencies. In their pricing model, Cong et al. (2021) propose a cryptocurrency fundamental-to-value ratio as the number of users over market capitalization. Further, they show that this ratio negatively predicts future cryptocurrency returns. Consistent with the literature on cryptocurrencies, there is no measure of book value for DeFi tokens. Corbet et al., (2021) and Maouchi et al., (2021) show that TVL can be used as a tool to monitor DeFi's success. TVL represents the total value allocated in a DeFi platform, and it can therefore be considered as (some of) its intrinsic value. The more people join and transact on DeFi platforms (translating in more funds and bigger TVL), the bigger the network will be. One of the Cong et al. (2021)'s main assumptions with regards to token valuation is based on the expected growth of the network as a result of the platform's increased productivity. Accordingly, we recognize TVL as a proxy for DeFi's 'book' value measure.

Before computing our valuation ratio, we investigate if the market values DeFi tokens based on their locked intrinsic value (TVL). For this, we simply plot the log market capitalization against the log. TVL, see Fig. 2. We found that there is a linear relationship between the two variables, meaning that DeFi tokens with high TVL have high market capitalization.

Furthermore, we construct the Book-to-Market ratio for the DeFi market by dividing the TVL by the Market Capitalization (MC). What we obtain is the TVL-to-Market ratio valuation, and we test its predictive power for DeFi token returns. For this analysis, we perform panel data regressions using the TVL-to-Market ratio. We first regress the DeFis returns (considering the DeFis for which we have the TVL information) on the TVL-to-Market ratio[8]. In a second step, we perform a panel data regression using only the 15 leading tokens (as we did in the previous sections). The reason behind this approach is to capture a bigger and more realistic picture of the DeFi market and not only results based on referring to the leading tokens. With these panel data regressions, we capture the impact of the TVL-to-Market ratio on the returns of DeFi.

---

[7] LINK, XTZ, MKR and LRC

[8] We are using TVL data in constructing this valuation ratio. As mentioned in the network effect section, we have TVL data for only 160 tokens, therefore, our TVL-To-Market ratio will be computed for 160 tokens as well.

Figure 2. The relationship between log. TVL and log. market value

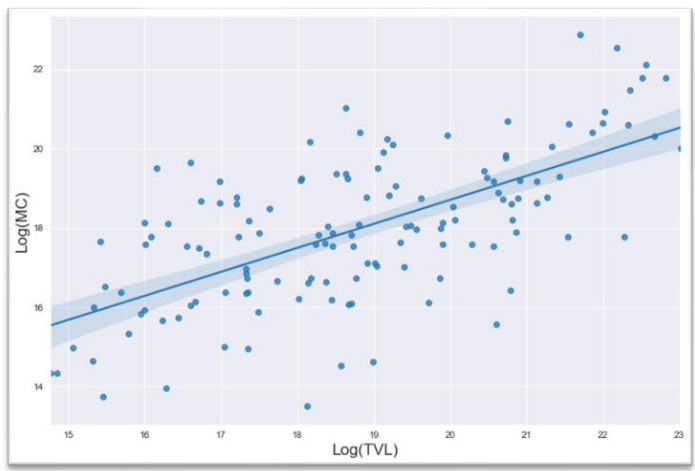

*In this graph, we can observe the linear relationship between DeFi's market capitalization and TVL. This suggests that the bigger the market capitalization is, the bigger the TVL.*

$$(16) \quad Rt\ DeFi\ Token_i(t) = \alpha_i(t) + \beta_{1,i}\frac{TVL}{MC}(t-1) + \beta_{2,i}\frac{TVL}{MC}(t-2) + \beta_{3,i}\frac{TVL}{MC}(t-3) + \beta_{4,i}\frac{TVL}{MC}(t-4) + \varepsilon_i(t)$$

The regression (16) tests the predictive power of our TVL-to-Market ratio on a DeFi token's returns. We do not make any initial assumptions about the ratio horizon of prediction, and we consider four different lags for the TVL-to-Market ratio. Results are shown in Table 15 and Table 16.

Table 15: Predicting power of the TVL-to-Market ratio for 160 DeFi tokens

|               | Parameter | Std. Err. | T-stat  | P-value | Lower CI | Upper CI |
|---------------|-----------|-----------|---------|---------|----------|----------|
| const         | 0.180     | 0.2941    | 0.6129  | 0.5400  | -0.3963  | 0.7568   |
| ValRatio(t-1) | 0.001     | 0.0036    | 0.1668  | 0.8676  | -0.0064  | 0.0076   |
| ValRatio(t-2) | **0.010**** | 0.0043    | 2.3257  | 0.0201  | 0.0016   | 0.0185   |
| ValRatio(t-3) | **-0.013**** | 0.0044    | -2.8435 | 0.0045  | -0.0211  | -0.0039  |
| ValRatio(t-4) | -0.001    | 0.0043    | -0.1437 | 0.8858  | -0.009   | 0.0078   |

*R2 is 0.27%. R2 Between is 0.32%. Panel regression of 160 DeFi returns against their Valuation Ratio = TVL/MarketCap. The data frequency is weekly. For this regression, we had 5289 observations. *** denotes significance levels based on the respective p-value (*:10%, **:5%, and ***:1%). The standard t-statistic value is shown in parentheses.*

Table 16: Predicting power of the TVL-to-Market ratio for 15 major DeFi tokens

|               | Parameter | Std. Err. | T-stat  | P-value | Lower CI | Upper CI |
|---------------|-----------|-----------|---------|---------|----------|----------|
| const         | 4.0165    | 1.0388    | 3.8664  | 0.0001  | 1.9762   | 6.0568   |
| ValRatio(t-1) | 0.6308    | 0.5607    | 1.125   | 0.2610  | -0.4704  | 1.732    |
| ValRatio(t-2) | -1.4726   | 1.1964    | -1.2309 | 0.2188  | -3.8223  | 0.8771   |
| ValRatio(t-3) | -1.2791   | 1.2731    | -1.0047 | 0.3155  | -3.7795  | 1.2213   |
| ValRatio(t-4) | 1.6722    | 1.0426    | 1.6039  | 0.1093  | -0.3754  | 3.7199   |

*R2 is 1.45%. R2 Between is 3%. Panel regression of major DeFi tokens against their Valuation Ratio = TVL/MarketCap. The data frequency is weekly. For this regression, we had 747 observations. The standard t-statistic value is shown in parentheses.*

As many researchers have shown that the book-to-market ratio predicts future returns on the stock market (Fama & French, 1992; Pontiff & Schall, 1998), we expect to reach the same conclusions about the DeFi market. Our results show that most of the DeFi token returns are positively (negatively) driven by the past increase in the TVL/MC ratio two (three) weeks ago. This indicates that when the value of a DeFi platform increases (relative to its market valuation), financial returns decrease in the following two weeks. On the other hand, Table 16 shows that leading DeFi's returns are not driven by our TVL/MC ratio. While our results are surprising, mainly thanks to the contrasting findings in week two and three, we think that they do not offer any obvious conclusions.

In the traditional markets, we often see studies (Ball et al., 2020; Pontiff & Schall, 1998) with Book-to-Market ratio that use monthly data. This is because the book value of a firm is renewed on a monthly basis. It is not the case for cryptocurrency and DeFi markets, where all the financial information is publicly available on Blockchain and constantly updated. For comparison purposes, we decided to use monthly data and investigate the ability of the TVL-to-market ratio to predict DeFi returns. The results are reported in appendix section D and are inconclusive.

## IV. Discussion and conclusion

In this work, we address the following global research question: 'What drives DeFi returns?'. Following Liu & Tsyvinski (2021)'s paper, which presents a comprehensive analysis of cryptocurrencies' returns, we performed an in-depth analysis of the determinants of the DeFi market returns. We consider several possible driving factors, such as: (1) the exposure of DeFi tokens to the cryptocurrency market, (2) the exposure to network variables, (3) the exposure to investor's attention and (4) the predictive power of TVL-to-Market valuation ratio.

Corbet et al., (2021) and Maouchi et al., (2021) show that DeFi tokens are distinct from the cryptocurrency asset class. Therefore, our first contribution is to compute a DeFi market benchmark based on the CRIX's methodology ('an index for Blockchain based currencies'). Considering that cryptocurrencies and DeFi tokens both run on Blockchain technology and belong to the crypto-market, our second contribution is to assess the exposure of DeFi token returns to the cryptocurrency market. Our results show that the cryptocurrency market strongly influences DeFi returns, fact confirmed by Corbet et al. (2021) and Yousaf & Yarovaya (2021).

The network effect in the crypto-market could be described as: a cryptocurrency's value and utility increases when more people join the network/ Blockchain. Vast literature (e.g., (Ante, 2020; Cong et al., 2021; Liu & Tsyvinski, 2021)) shows that the cryptocurrency market is highly impacted by its network effect. Therefore, our third contribution is to assess if the same holds for DeFi tokens. We thus have measured the network effect in the DeFi market by using three proxies: the transaction count, unique addresses count, and TVL. If the first two variables have been previously used in the cryptocurrency-related literature, we are, as far as we know, the first ones to use the TVL. This represents the fourth contribution of this paper. TVL is a unique variable characteristic of the DeFi market. It reflects the amount of funds committed to DeFi platforms, and it is an indicator of market growth and success. While all three network variables seem to have an important impact on DeFis returns, the transactions and TVL seem to be the most significant ones. Here we also show that DeFi returns are negatively impacted by the transaction and address growth. This could be explained by the presence of high volatility and large financial bubbles in the DeFi market (Maouchi et al., 2021). Similar findings are presented in (Hau et al., 2021), which shows that cryptocurrency returns can be positively (negatively) influenced by transaction activity if the market state is bullish (bearish).

A vast literature has shown the importance of investor attention in asset pricing for both traditional markets and cryptocurrencies (Andrei & Hasler, 2015; Ciaian et al., 2015; Lin, 2020; Liu et al., 2022). Similarly, the fifth contribution of this paper is to assess whether investor's attention has an impact on the DeFi market as well. We construct proxies for investor attention with Google searches for the "decentralized finance" and "DeFi" terms. Our results show that high investor attention, proxied by "DeFi" term search, predicts high future returns over one to three-weeks horizons. Google searches for "decentralized finance" term are negatively impacting DeFis, meaning that in this case, high investor attention predicts lower future returns over one to three weeks. An interesting fact is that the tokens impacted by investor's attention, are also the ones strongly impacted by cryptocurrencies, with a predictability statistically significant (high R squared). Overall, our analysis shows that Google searches have a significant and strong impact on DeFi returns.

Motivated by the existing literature on the cryptocurrency market and studies of other financial markets, we investigate if DeFi returns can be predicted by their 'book-to-market' ratio. As there is no standard 'book' value for DeFi tokens, we have constructed a Book-to-Market ratio for the DeFi market by dividing the TVL by the Market Capitalization (MC). This represents the sixth contribution of this paper. TVL represents the total value allocated in a DeFi platform, and it can therefore be considered as (some of) its intrinsic value. The results obtained do not offer any obvious answers. Hence, we conclude that there is not enough evidence to support our assumption that the TVL-to-Market ratio contains information about future DeFi returns.

Overall, our empirical study shows that the impact of the cryptocurrency market on DeFi returns is stronger than any other driver considered in this analysis and provides superior explanatory power. A possible limitation for this paper is the restricted information available for network data: transaction and address count.

As a future avenue for investigation, it would be interesting to see how network factors impact DeFi returns during different market states. Another idea is to group and assess DeFi returns based on their volatility level. Further evidence concerning DeFis' price efficiency would greatly complement the findings of this study. There is an obvious need for research on the NFT market as well. Dowling (2022) and Karim et al. (2022) reveal that NFTs are a distinct asset class from conventional cryptocurrencies; therefore, it would be interesting to construct a benchmark for this market and analyze the proprieties of NFTs prices.

# Appendix

### A. A brief DeFi market description

Cryptocurrencies aim to replace or offer alternative payment tools and money, while DeFi platforms seek to revolutionize the financial system as a whole. Like bitcoin, DeFi platforms run on Blockchains (distributed ledger) technology, onto which decentralized applications (based on smart contracts) are added (Popescu, 2020). DeFi's applications provide financial services that rely on cryptocurrencies and crypto tokens. Their goal is to provide a digital alternative to traditional banking, exchange, and investment services (Anker-Sørensen & Zetzsche, 2021). With decentralized applications (dApps) deployed on the Blockchain, DeFi can bring numerous benefits, among which reduced operational costs, borderless financial service access, and improved privacy. Like cryptocurrencies, DeFi uses public Blockchains that make it accessible to anyone with an internet connection. Decentralized finance platforms are often compared to puzzles or Lego mainly because these tools are complex and use multiple technological layers (Katona, 2021; Popescu, 2020; Schäfer, 2021). Anyone interested in developing a new decentralized finance solution can get the source code of the existing platforms and create innovations while combining parts of the current applications.

It is important to mention that regardless of their scope, DeFi platforms are very different from one another in terms of both development and operation. While constructed around similar principles as traditional finance, decentralized finance works in substantially different ways (Aramonte et al., 2021). For example, loans in the DeFi world are not always using physical collateral but rather some digital assets (e.g., Non-Fungible Tokens – NFT, bitcoins, ether) that, once deposed, will allow the borrowing of another digital coin. In this scenario, the user lending capital earns interest without the intervention of any central authority, and the borrower can further invest (without intermediaries) its funds in other services such as trading or portfolios (Corbet et al., 2021). As describing the whole business model of DeFi platforms is out of our scope, a detailed description can be found in (Harwick & Caton, 2020; Ramos & Zanko, 2021; Schär, 2021; Stepanova & Eriņš, 2021; Zetzsche et al., 2020; Zumwalde et al., 2021).

From a technological perspective, DeFi tokens (abr. DeFis) and cryptocurrencies are similar in the sense that both are based on Blockchain technology and implement decentralized (automatic) management.

Compared to cryptocurrencies' functions (money-like) (Baur & Dimpfl, 2021; Hazlett & Luther, 2020), DeFi tokens resemble more to ICO tokens. ICO tokens can fulfill multiple roles, more specifically, they can be used to obtain products or services, can be traded on a platform (secondary market), and/or could be held to earn a profit (Le Moign, 2019). As DeFi platforms can perform most of the things that banks do — lend, trade assets, earn interest, buy insurance, borrow, trade derivatives, and more (Coinbase, n.d.), these activities are supported by complex tokens with different functions depending on the platform's needs. For example, DeFi platforms can have transactional tokens (e.g., stablecoins: Dai, TUSD, USDC, WBTC) that facilitate fund transfers across platforms (Aramonte et al., 2021), governance tokens (e.g., MKR, COMP, YFI) that enable users to take part in the platform development, resembling with the common stock, utility tokens are used in the same way as in ICOs, to obtain access to the platform's services (in-App payment '*currencies*'), liquidity provider (LP) tokens are used as a reward for the users contributing to the liquidity of a DEX (decentralized exchange), and collateral tokens are used on lending platforms, a practice similar to the bank loans (e.g., stablecoins, LP, ETH, NFTs) (MakerDAO.com, 2021). Despite their distinct nature, ICO tokens have been often studied together with other digital assets, first considering the relationship between Blockchain tokens and cryptocurrencies, and second, considering that they all belong to the same crypto-market (Fahlenbrach & Frattaroli, 2021; Howell et al., 2020; A. S. Hu et al., 2019; Lyandres et al., 2018; Maouchi et al., 2021).

An important variable used in this paper is the **t**otal **v**alue **l**ocked (abr. TVL). TVL is a unique variable characteristic of the DeFi market and refers to the amount of funds attached to a DeFi project. More specifically, if we take the example of lending platforms, TVL is the amount put into DeFi projects as collateral for the loans taken. To compute the TVL, we multiply the amount of crypto-assets staked as collateral on the Blockchain by their current price.

## B. DeFiX computation

To the best of our knowledge, the development of the entire DeFi market has not yet been studied, and only samples have been taken into account (Corbet et al., 2021; Karim et al., 2022; Stepanova & Eriņš, 2021; Yousaf et al., 2022). We contribute to this area of research and design "DeFiX", a market index (benchmark) that will enable each interested party to study the performance of the DeFi market as a whole.

Similar to the CRIX index, which was created out of the need to assess the emerging market of cryptocurrencies accurately, DeFiX aims to represent a new asset class for which public interest has arisen: DeFi tokens. According to its name (Trimborn & Härdle, 2018), CRIX is '*an Index for Blockchain-based currencies*'. Considering the fact that cryptocurrencies and DeFi tokens run on Blockchain technology and belong to the same crypto-market, we find it appropriate to use CRIX index methodology and a variant of its original code[9] to construct DeFiX.

CRIX formula is a derivation of Laspeyres index: $Index = \frac{\sum_i P_{it} Q_{i0}}{\sum_i P_{i0} Q_{i0}}$, where $P_{it}$ is the price of asset $i$ at time $t$, and $Q_{i0}$ is the quantity of asset $i$ at time $0$. Laspeyres statistic represents a consumer price index and has been developed to measure the price change of the basket of goods and services consumed.

Adjusting the index for capital markets makes the quantity $Q_{i0}$ a measure for the number of shares of the asset $i$, which is multiplied with its corresponding price, resulting in the market capitalization. Therefore, the index components (stocks) are weighted by their market capitalizations. As markets change all the time, it is important that the index contains the most appropriate market representatives as its members. Hence, the components of the index must change in tone with the market state. To make this possible, there are several adjustments in the CRIX formula (Trimborn & Härdle, 2018):

$$(B.1.) \; CRIX_t \, (k, \beta) = \frac{\sum_i^k \beta_{i,t_l^-} P_{it} Q_{i,t_l^-}}{Divisor(k)_{t_l^-}}$$

Where $k$ is the number of constituents, $Q_{i_l}$ represents the amount (quantity) of coin $i$ at time $t$, $P_{i_l}$ is the price of coin $i$ at time $t$, $\beta_{i,t_l^-}$ represents the adjustment factor of cryptocurrency $i$ at time $t_l^-$, $l$ indicates that this is the $l$-th adjustment factor, and $t_l^-$ is the last time point when $Q_{i,t_l^-}$, $Divisor(k)_{t_l^-}$ and $\beta_{i,t_l^-}$ have been updated. The denominator has been replaced by a divisor which is specific to each index (Trimborn & Härdle, 2018). Its formula is:

$$(B.2.) \; Divisor \, (k, \beta)_0 = \frac{\sum_{i-1}^k \beta_{i0} \times P_{i0} \times Q_{i0}}{Base \; value}$$

The divisor is changing every time there is a change in the number of components (k) and ensures that the adjustments are stable. The starting value of the CRIX Index is 1000 (Trimborn & Härdle, 2018).

The selection criteria for cryptocurrencies before being included in the index is made based on the Akaike Information Criterion (AIC), meaning that only the most representative cryptocurrencies of the market will be taken into consideration. More specifically, cryptocurrencies must have sufficient price movements, meaning that they are actively traded and are among the largest representatives by market capitalization (over 1,000,000 USD). As a common practice for the stock market indexes, CRIX members will always be a multiple of five (Trimborn & Härdle, 2018).

DeFiX benchmark is computed using the exact same steps as described above in the CRIX methodology. We have removed the DeFi tokens with less than 1,000,000 USD market capitalization. Then, we ran the CRIX code with our DeFi data and built a value-weighted market index of the DeFi market as a whole, comprising 95 components.

---

[9] Retrieved from www.quantlet.de

DeFiX can be found and downloaded from: https://hpac.imag.fr/cryptotracker/dashboards/defix.html.

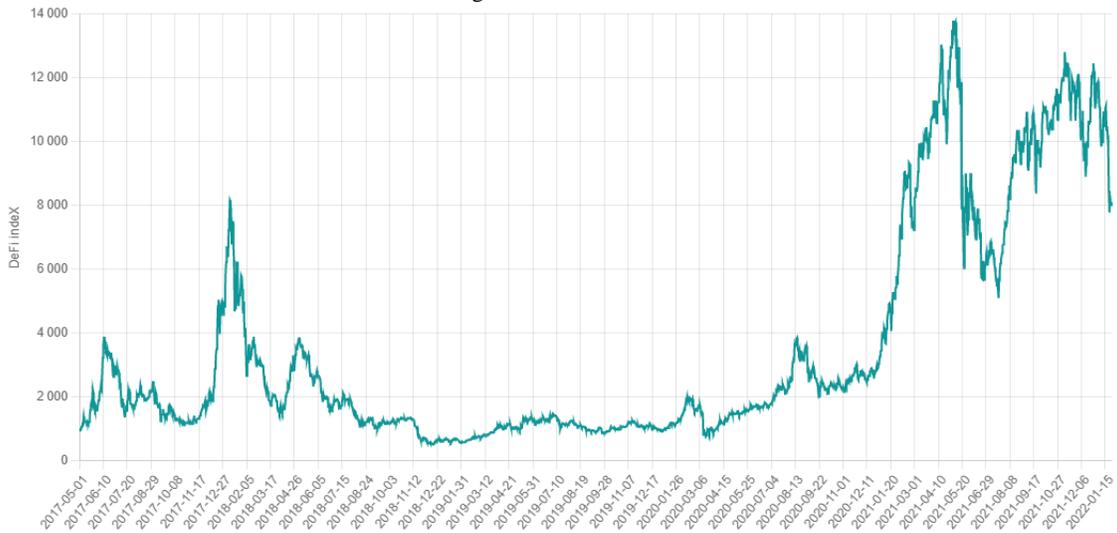

Fig. B.1. DeFiX index

## C. Data description

Table C.1. Financial data

| (Token) Symbol | Description / DeFi platform | Daily data observations | Time span |
|---|---|---|---|
| **CRIX** | Index for Cryptocurrency market | 1009 | 19 March 2018 to 27 January 2022 |
| **BTC** | Cryptocurrency | 3176 | 28 April 2013 to 9 January 2022 |
| **ETH** | Cryptocurrency | 2346 | 7 August 2015 to 9 January 2022 |
| **DeFiX** | Index for DeFi market | 1730 | 1 May 2017 to 25 January 2022 |
| **LUNA** | Terra | 916 | 26 July 2019 to 26 January 2022 |
| **AVAX** | Avalanche | 494 | 13 July 2020 to 26 January 2022 |
| **WBTC** | Wrapped Bitcoin | 1093 | 30 January 2019 to 26 January 2022 |
| **DAI** | MakerDAO | 797 | 22 November 2019 to 26 January 2022 |
| **LINK** | Chainlink | 1590 | 20 September 2017 to 26 January 2022 |
| **UNI** | Uniswap | 497 | 17 September 2020 to 26 January 2022 |
| **FTM** | Fantom | 1185 | 30 October 2018 to 26 January 2022 |
| **XTZ** | Tezos | 1578 | 2 October 2017 to 26 January 2022 |
| **AAVE** | Aave | 482 | 2 October 2020 to 26 January 2022 |
| **GRT** | The Graph | 406 | 17 December 2020 to 26 January 2022 |
| **CAKE** | PancakeSwap | 485 | 29 September 2020 to 26 January 2022 |
| **MKR** | Maker | 1546 | 29 January 2017 to 26 January 2022 |
| **RUNE** | THORChain | 919 | 23 July 2019 to 26 January 2022 |
| **CRV** | Curve DAO Token | 531 | 14 August 2020 to 26 January 2022 |

| | | | | |
|---|---|---|---|---|
| LRC | Loopring | 1611 | 30 August 2017 to 26 January 2022 | |

This table describes the data for financial variables used: CRIX index and the two main cryptocurrencies (BTC and ETH), DeFiX index and the 15 main DeFi tokens.

Table C.2. Non-financial data

| Description | Daily data observations | Time span |
|---|---|---|
| **TVL** | 1009 | 19 March 2018 to 27 January 2022 |
| **Transactions count** | 26735 | 9 April 2016 to 3 March 2022 |
| **Addresses count** | 26735 | 9 April 2016 to 3 March 2022 |
| **Google search** "*Decentralized finance*" | 254 (weekly) | 5 March 2017 to 9 January 2022 |
| **Google search "*DeFi*"** | 254 (weekly) | 5 March 2017 to 9 January 2022 |

This table describes the data for non-financial variables used: the network proxies (TVL, transactions and address count) and the proxies for investor attention (google search terms "Decentralized finance", "DeFi"). The data downloaded for investor attention from Google has weekly frequency.

### D. Regressing DeFi tokens returns against their valuation ratio

$$(D.1.) \; Rt \, DeFi \, Token_i(t) = \alpha_i(t) + \beta_{1,i} \frac{TVL}{MC}(t-1) + \beta_{2,i} \frac{TVL}{MC}(t-2) + \beta_{3,i} \frac{TVL}{MC}(t-3) + \beta_{4,i} \frac{TVL}{MC}(t-4) + \varepsilon_i(t)$$

Table D.1.: Predicting power of the TVL-to-Market ratio for 160 DeFi tokens

| | Parameter | Std. Err. | T-stat | P-value | Lower CI | Upper CI |
|---|---|---|---|---|---|---|
| *const* | **3.625**\*\* | 1.5289 | 2.371 | 0.0179 | 0.6246 | 6.6255 |
| *ValRatio(t-1)* | -0.001 | 0.0155 | -0.065 | 0.9482 | -0.0313 | 0.0293 |
| ValRatio(t-2) | **-0.3715**\*\* | 0.1657 | -2.2416 | 0.0252 | -0.6968 | -0.0463 |
| *ValRatio(t-3)* | **0.6248**\*\*\* | 0.2341 | 2.6683 | 0.0078 | 0.1653 | 1.0842 |
| *ValRatio(t-4)* | **-0.3772**\*\* | 0.1558 | -2.4216 | 0.0156 | -0.6829 | -0.0715 |

R2 is 1.45%. R2 Between is 3.36%. Panel regression of 160 DeFi returns against their Valuation Ratio = TVL/MarketCap. The data frequency is monthly. For this regression, we had 981 observations. \*\*\* denotes significance levels based on the respective p-value (\*:10%, \*\*:5%, and \*\*\*:1%). The standard t-statistic value is shown in parentheses.

Table D.2.: Predicting power of the TVL-to-Market ratio for major DeFi tokens

| | Parameter | Std. Err. | T-stat | P-value | Lower CI | Upper CI |
|---|---|---|---|---|---|---|
| *const* | **21.012**\*\*\* | 5.0462 | 4.1638 | 0.0001 | 11.011 | 31.012 |
| *ValRatio(t-1)* | 3.1015 | 3.268 | 0.949 | 0.3447 | -3.3749 | 9.5778 |
| *ValRatio(t-2)* | **-8.8164**\* | 4.6141 | -1.9108 | 0.0586 | -17.96 | 0.3276 |
| *ValRatio(t-3)* | -2.5913 | 5.3891 | -0.4808 | 0.6316 | -13.271 | 8.0886 |
| *ValRatio(t-4)* | 5.9337 | 4.6433 | 1.2779 | 0.204 | -3.2682 | 15.13 |

R2 is 10.18%. R2 Between is 23.21%. Panel regression of major DeFi tokens against their Valuation Ratio = TVL/MarketCap. The data frequency is monthly. For this regression, we had 148 observations. \*\*\* denotes significance levels based on the respective p-value (\*:10%, \*\*:5%, and \*\*\*:1%). The standard t-statistic value is shown in parentheses.

By looking at our results, we find that most of the DeFi token returns are strongly negatively (and significantly) driven by the past increase in the TVL/MC ratio (two and four months ago). This indicates that when the value of a DeFi platform decreases (relative to its market valuation), financial returns increase in the following two and four months. At the same time, if the ratio goes down (market capitalization increases or TVL decreases), it means that the market overvalues the DeFi, translating in an increase (correction) of the DeFi returns two and four months after. Consistent with this result, our regression on leading tokens shows as well that there is a negative and significant relationship between DeFi returns and the two previous months' TVL/MC ratio.

A surprising result is that contrary to the current results, DeFi token returns seem to be strongly and positively driven by three previous months' TVL/MC ratio. Therefore, we conclude that our analysis provides inconclusive evidence.